# Untangling the Dueling Expert Witnesses: Comparing Ensemble Methods in Pennsylvania's Redistricting Plans

Paul Dingus[1], Catherine Zhu[2] & Constantine Gonatas[3]

[1] Harvard Kennedy School
[2] NERA Economic Consulting
[3] CPG Advisors; correspondence to cpgonatas@cpg-advisors.net

Abstract

Ensembles of random legislative districts are a valuable tool for assessing whether a proposed district plan is an outlier or gerrymander. Expert witnesses have presented these in litigation using various methods, and unsurprisingly, they often disagree.

Recent open source methods now permit independent validation of expert witness testimony. Here, we compare ensembles for the Pennsylvania House and Congressional districts calculated using "Redist" and "Gerrychain" further incorporating constraints restricting county and municipal boundary splitting, as required by Pennsylvania for legal plans. We compare results to expert witness testimony submitted by Republican and Democratic parties. We confirm some of the testimony but could not reproduce all of it, struggling with metrics based on a heuristic "sum of votes index" rather than a straightforward average of metrics across multiple elections. We recommend against relying on analytics based on summing votes from multiple elections to create vote indices and derivative metrics as these are inherently poorly behaved. To promote transparency, we recommend that where possible, expert witness testimony be based solely on publicly available election data as opposed to proprietary data closely held by political parties.

Introduction

Partisan majorities in state legislatures have gerrymandered legislative and congressional districts for decades, on a bi-partisan basis. In the current redistricting cycle, the Wisconsin,

Georgia, Ohio, Texas, Florida, Illinois, and Maryland legislatures enacted district plans giving the

political parties in control advantages that range from a quantitative edge of a few % seats out

of proportion to vote share. In general, a legislature unchecked by any law or countervailing

authority can draw plans that lock the incumbent party in place nearly regardless of election

results. For example, the Wisconsin state legislative plans returned 62% Republican majorities

in the General Assembly and State Senate notwithstanding the state being nearly evenly

divided, with a slight pro-Democratic voteshare. According to Dave's Redistricting, Democrats

would need to achieve an additional 4.2% over 50% voteshare to reach parity in the General

Assembly[4]. The 2021 Princeton Gerrymandering Map-off[5] featured "stealth gerrymanders"

(Cervas and Grofman 2020) where a party could sustain majorities in Legislatures despite a minority voteshare with compact districts and moderate numbers of county splits that would not attract attention as egregious gerrymanders.

Although Federal courts have closed claims of partisan unfairness in districting (*Rucho vs*

*Common Cause* (2019) S. Court 18-422, stating claims of partisan unfairness were

"injusticiable"), State courts have considered these claims based on State Constitutions,

including ballot propositions creating provisions that bar favoring or disfavoring a political party. In Ohio, voters approved Issue 1 in 2018, creating a process whereby a "plan cannot unduly favor or disfavor a political party or incumbents. In Michigan, a citizens group "Voters not Politicians" gathered enough strength to pass proposal 18-2 in 2018 including language "Districts shall not provide a disproportionate advantage to any political party." Years earlier, in 2010, Florida voters passed Amendments 5 & 6 barring districts "drawn to favor or disfavor an incumbent or political party." In New York State, a legislatively referred ballot proposition amended the State Constitution, creating a redistricting commission and enable a process that would prohibit "drawing districts to favor or disfavor any incumbents… or political parties,[6]" whose criteria were invoked by the NY Supreme Court to bar a legislatively drawn plan that sought an end-run around the newly constituted Independent Redistricting Commission.

---

[4] Davesredistricting.org
[5] https://gerrymander.princeton.edu/press-release-map-contest

Exiting State Constitutional provisions, while not as explicit as the above ballot questions, have permitted State Courts to invalidate legislatively drawn plans. The North Carolina Constitution of 1971 states "Elections shall be free." The Pennsylvania State Constitution, Article I §5, states "Elections shall be free and equal." The respective State Supreme Courts have interpreted these provisions, albeit along party lines, to bar gerrymandered plans[7]. Expert witness testimony challenging the legislatively enacted plans sought to show those were outliers as compared to ensembles of random plans (Mattingly 2019) while those supporting them sought to show they were close to the median of a distribution.

---

[6] https://ballotpedia.org/New_York_Redistricting_Commission_Amendment,_Proposal_1_(2014)
Bipartisan sponsorship from Sheldon Silver (House) and Dean Skelos (Senate)
[7] for NC, NC League of Conservation Voters vs. Hall, 21 CVS 500085; for PA, League of Women Voters vs. Commonwealth of Pennsylvania, No 261 MD 2017 and No 159 MD 2017, dueling expert witness reports by Chen, J (for Democrats) and McCarty, N (for Republicans): https://www.pubintlaw.org/wp-content/uploads/2017/06/Expert-Report-Jowei-Chen.pdf and
https://www.brennancenter.org/sites/default/files/audio/LWV_v_PA_Expert_Report_NolanMcCarty.pdf

Methods for simulating ensembles have been made available recently as open source models. These are valuable tools for validating expert testimony, often composed under time pressure and without full disclosure of methods or simulation conditions. Historically, expert testimony featuring ensemble simulations did not usually track or constrain county splits so sample plans in ensembles would not resemble legal plans. The open source simulation methods can incorporate these constraints as well as constraints on compactness, population deviation and so on. While it is not important for ensemble plans to adhere to legal requirements exactly – for example, it is pointless to require Congressional models to have +/-1 person population deviation as such a limitation would not give a more accurate measure of partisanship metrics, it is helpful for ensemble elements to have the look and feel of actual districts. In this paper we compare simulations using multiple methods in Redist (Harvard) and Gerrychain (Tufts) to gain insight on differences. Then we compare our results to actual recent expert witness testimony filed in Pennsylvania litigation in an attempt to validate the "battle of the experts."

Historically, much of the emphasis on ensembles has been showing that an enacted plan was an outlier compared to a distribution. Research has examined deviations between median and mean as a metric of partisan gerrymandering (McDonald and Best 2015). Some recent expert witness reports often claim deviations in a proposed plan metric from an ensemble mean always implies bias. Yet the political geography of a State may result in a distribution that has a built-in bias towards one party. Such "unintentional gerrymandering" can result in anomalous

outcomes, where a minority of voters can consistently achieve a majority in the state

legislature. This is often a result where Democrats win cities with overwhelming majorities

while being too dispersed elsewhere to gain more than a handful of suburban seats (Rodden

2019).

Some argue that leveling plans for partisan bias implies a right to proportional representation,

something not guaranteed by the US Constitution[8]. Yet it is *not* demanding proportional representation to seek symmetry in partisan outcomes (Grofman and King 2007) nor ask that the "democratic ideal that the party attracting about 50% of the popular vote, also ought to be winning about 50% of the contested seats.[9]" Simple majoritarian rule differs from a proportional representation claim a party winning x% of the vote is entitled to x% of the seats.

With respect to whether the mean or median of an ensemble is the "right" level of partisanship in a map, Becker et al. (2021) argue ensembles are best for determining

"Normal range, not ideal. We advocate using redistricting ensembles to learn a normal range for metrics and measures under the constraints of a set of stated redistricting rules and priorities. Ensembles allow us to justify statements such as Plan X is an outlier in its partisan lean, taking all relevant rules into account. While talking about normal ranges and outliers, we should avoid the temptation to valorize the top of the bell curve (or its center of mass, or any other value) as an ideal. By analogy, we can talk about people who are unusually tall or short without believing that any height is most desirable or ideal. If the 50[th] percentile height for American women is 5'4" and the 99th percentile height is 5'10", we can conclude that a woman who is six feet tall is unusual, and we can look for reasons (family history, diet, and so on) to explain her height. But it would be quite strange to decide that a woman who is 5'4" is a better" height than one who is 5'5."

We present results from ensemble distributions calculated several different ways using public source methods (Gerrychain and Redist with various algorithms within) then compare to expert witness testimony as best as possible to validate the methods and testimony. Thus we hope to clarify some of the confusion resulting from dueling experts and methods! We focus on recent Pennsylvania cases for Statehouse and Congress where testimony was presented using ensembles with contradictory results, but approximately enough detail we could attempt to reproduce their results. Pennsylvania has a Constitutional requirement not only restricting county splits, but splits of municipal boundaries, making ensemble analysis particularly challenging[10].

The more constrained an ensemble the more difficult it is for a random map generator to create proposed plans without violations. Because many PA municipalities are small – and there are 2570 of them as compared to 67 counties – the more proposed plans an algorithm

---

[10] Pennsylvania Constitution II § 16 "Unless absolutely necessary no county, city, ... shall be divided in forming either a senatorial or representative district." LWV vs PA No 261 MD 2017 and No 159 MD 2017 construed this restriction to Congressional districts

generates are rejected for boundary splitting rules. Our methods only consider whole precincts, without splitting any into census blocks, raising the possibility that members of an ensemble calculated with a Markov chain could be just a few unrepresentative maps with the same precincts shuffled from one to another. We assess the validity of an ensemble approach, comparing Markov chain simulations to results of methods that calculate each ensemble independently (where possible), where each ensemble member is inherently uncorrelated.

Overview of Algorithms

We performed simulations of PA House and Congressional Districts (HD and CD respectively) using a combination of algorithms from Redist and Gerrychain to compare results and performance between each other,

Gerrychain[11] is a product of the Tufts Metric Geometry Group. We used its python-based "Recom" (DeFord, Duchin and Solomon 2021) and random tree algorithms to generate quasi-random ensembles of maps to sample the space of possible districting plans. The Recom algorithm acts on a Markov chain of successive district plans translating a chain step to a subsequent step by merging two adjacent districts, drawing a spanning tree covering the combined districts. Each step randomnizes the plan incrementally, analogous to shuffling a deck of cards by removing a few cards at a time and replacing them in random places in the deck.

---

[11] Mggg.org; https://github.com/mggg/GerryChain

For example, in Figure 1 we show an initial state for a Congressional map in Iowa followed by subsequent Markov chain step where Recom acts on the initial state once. Here, Recom has combined the two districts in the Northern part of the State then re-divided them to find a new boundary, subject to the equal population and contiguity constraints while leaving the Southern districts unchanged.

Figure 1: Recom Algorithm Step Applied to Iowa Congressional Plan

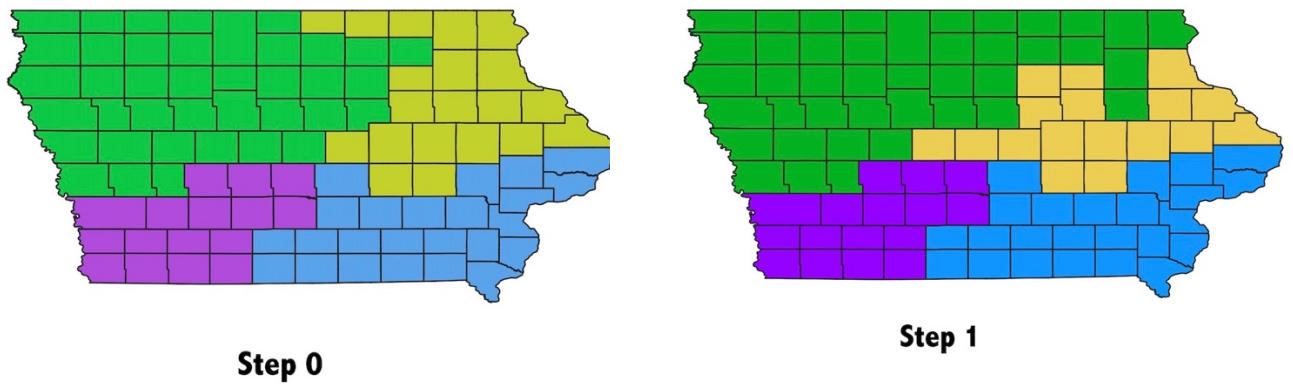

For a plan with only a few districts, it only takes a small number of Recom steps in a Markov chain to completely randomnize a plan, obtaining an ensemble of uncorrelated states that one can hope provide an unbiased sample of the enormous space of possible plans. We use the autocorrelation function of a plan metric, such as democratic seats won, to infer state mixing. In Figure 2 we show this metric's autocorrelation for an ensemble of Pennsylvania Congressional plans, showing a rapid decline from unity. It declines to near zero within 50 steps and thereafter stays low with some oscillation about zero, indicating good state mixing and suggesting a representative sampling of the space of all maps.

Figure 2: Autocorrelation of a Metric in a Markov Chain Ensemble for PA Congressional Districts

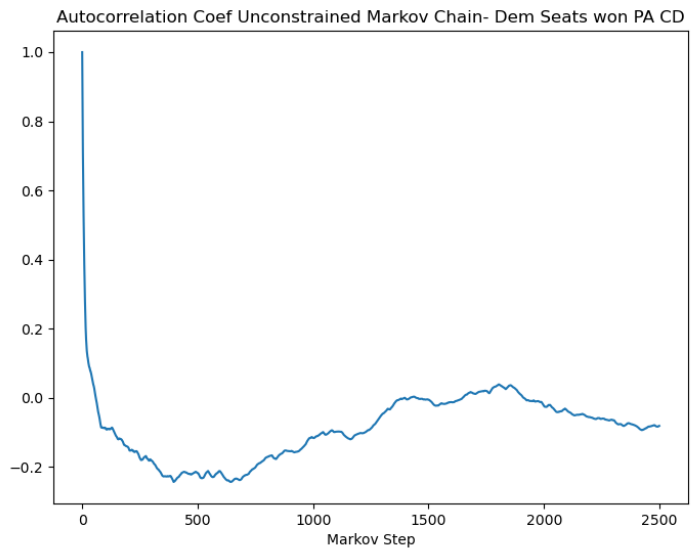

Within Gerrychain, another technique for generating maps is creating a new partition using a random tree. This is far slower than evolving a Markov chain stepwise but has the advantage that each member of a map collection created from n random trees is inherently uncorrelated. Because of its inefficiency it is not possible to impose many constraints within this method but it is useful as a benchmark for a random ensemble of maps with a simple constraints (eg. population balance, contiguity).

Redist, introduced by Fifield et al. 2020 is an R-based package containing several methods to perform Monte Carlo generation of random district plans. One method, the "merge-split" algorithm is an R implementation of Gerrychain's Recom proposal for generating subsequent Markov chain steps from an initial plan configuration.

Redist introduces a novel algorithm, "Sequential Monte Carlo" (SMC, McCartan and Imai 2022) that generates new map proposals ab initio for each plan instead of using a Markov chain evolution. This is an alternate approach from merge-split or Gerrychain's Recom. Here, an initial district is typically drawn somewhere at random while minimally splitting county boundaries. Additional districts are attached to the previously drawn districts one by one while minimizing county splits (or whatever specified administrative boundaries). Each new plan starts drawing from scratch thus is uncorrelated to prior plans. We illustrate this in Figure 3 (from McCartan & Imai 2022).

It not obvious how SMC avoids painting itself into a corner. Near the end of a run there are only a few districts left to paint so it might not be possible to draw contiguous or compact districts in the unassigned space remaining. While Redist-SMC worked well for PA Congressional districts, it failed to converge for PA House districts. In expert witness testimony[12], Kosuke Imai overcame this problem by dividing Pennsylvania into 5 dense regions, using SMC to apportion districts for those separately, then joining with the remainder of the State.

---

[12] https://www.pacourts.us/Storage/media/pdfs/20220314/163459-march11,2022-intervenormcclinton'sbrief.pdf
brief of Johanna McClinton, appendix D; filed under docket 11 MM 2022 PA Supreme Court

Figure 3: Redist SMC Illustrated for IA Congressional Districts (from McCartan & Imai 2022)

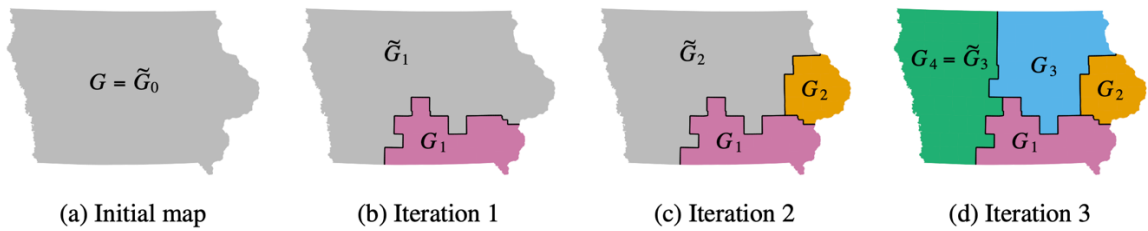

Redist-SMC ensemble elements are uncorrelated with each other, even when the ensemble has many constraints. Redist merge-split for an unconstrained ensemble (no county or municipal split constraints) shows an autocorrelation function (Figure 4) that decays very quickly with markov chain step, similar to Gerrychain.

Figure 4: Autocorrelation for Redist merge-split:

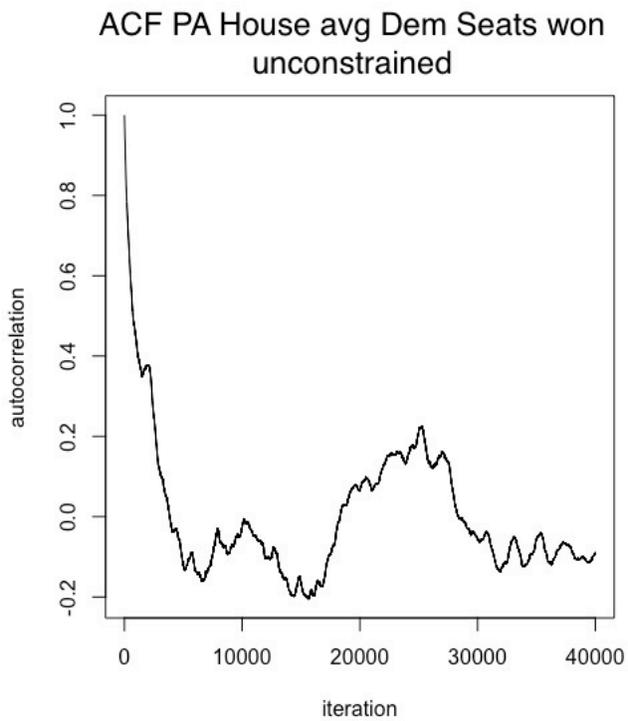

Performance comparisons:

We benchmarked Redist vs Gerrychain, on an Apple M1 processor (3.2 GHz), showing results in Table 1. There is a remarkable speed difference between the two packages because Redist's computationally intensive pieces are written in C. Gerrychain is in python (with an experimental version in Julia): much slower but easier to understand and customize.

Reading in map data for Pennsylvania and converting shapefiles into a connected graph was 2.8x faster in Redist. Running a basic, unconstrained merge-split chain (1000 chain iterations) was 30x faster in Redist. With a constrained chain run (county & municipal split constraints), Redist merge-split was even more efficient than Gerrychain, accepting 24% of proposed Markov steps, and only a few seconds faster than the unconstrained run. For the inefficient county & municipal split rejection criteria we introduced in Gerrychain, simply rejecting any proposals exceeding set limits for splits, Gerrychain took nearly 32x longer than an unconstrained run, and nearly 400x longer than the constrained Redist run.

Unconstrained Redist SMC was substantially slower than Redist merge-split but counter-intuitively, became faster with additional constraints. It had a substantial advantage over merge-split of exploring the space of feasible maps without being confined to subsequent steps of a markov chain, risking potential auto-correlation problems. Constrained Redist SMC is about 30x faster than a constrained Gerrychain. Unfortunately, SMC failed to converge in runs for the PA House even though it performed well for PA Congress (CD).

Table 1: Gerrychain and Redist Speed Benchmarking (PA Congressional District Simulations)

| PA CD | read-in/ graph | sec for 1000 iterations |
|---|---|---|
| gerrychain (no constraints) | 50 | 475 |
| gerrychain + 2 constraints | - | 6829 |
| gerrychain random tree | - | 10142 |
| redist merge-split | 18 | 15 |
| redist merge-split+ 1 constraint | - | 18 |
| redist merge-split + 2 constraints | - | 415 |
| redist smc | 18 | 558 |
| redist smc 1 constraint | - | 250 |
| redist smc 2 constraints | - | 325 |

Election Data

We have access to 2012-2020 precinct election data for Governor, US Senator, President and Attorney General but unfortunately lack precinct data for State Auditor and Treasurer that were included in the expert witness testimony. For comparability to one of the expert witnesses (Michael Barber, for PA House Republicans) we also tally a "vote index."

As opposed to our default method of averaging metrics over their value per each election, the vote index alternative, popularized by voter data expert Thomas Hofeller, is a heuristic obtained by summing votes over all elections in a sample and only thereafter computing metrics. That is, if there are three elections in a sample and a candidate won 10 voters in a precinct in the first election, 15 in the second and 11 votes in the third, the vote index for that precinct would be 36 total. Thus measuring by vote index overweights contests held in presidential election years and de-weights down-ballot contests where voters cast blank entries more often than for the top of the ballot. We show statewide voteshares by individual election and averages over 2012-20 and 2016-20 periods by election and for the vote index in Table 2.

Table 2: Statewide Partisan Voteshares

a. Voteshares by election

| Statewide Dem Voteshare: | | votes D | votes R | 2-party D% |
|---|---|---|---|---|
| *Auditor Gl 2012* | | *2,729,565* | *2,548,767* | *51.71%* |
| *Treasurer 2012* | | *2,872,344* | *2,405,654* | *54.42%* |
| Atty general 2012 | | 3,125,557 | 2,313,506 | 57.46% |
| Senate 2012 | | 3,021,364 | 2,509,132 | 54.63% |
| President 2012 | | 2,990,274 | 2,680,434 | 52.73% |
| Governor 2014 | | 1,920,355 | 1,575,511 | 54.93% |
| *Auditor Gl 2016* | | *2,958,818* | *2,667,318* | *52.59%* |
| *Treasurer 2016* | | *2,991,404* | *2,610,811* | *53.40%* |
| President 2016 | | 2,926,441 | 2,970,733 | 49.62% |
| Atty general 2016 | | 3,057,010 | 2,891,325 | 51.39% |
| Senate 2016 | | 2,865,012 | 2,951,702 | 49.25% |
| Senate 2018 | | 2,792,437 | 2,134,848 | 56.67% |
| Governor 2018 | | 2,895,652 | 2,039,882 | 58.67% |
| *Auditor Gl 2020* | | *3,129,131* | *3,338,009* | *48.39%* |
| *Treasurer 2020* | | *3,239,331* | *3,291,877* | *49.60%* |
| Atty general 2020 | | 3,461,472 | 3,153,831 | 52.33% |
| President 2020 | | 3,458,229 | 3,377,674 | 50.59% |

b. Averages over 2012-20 and 2016-20 elections

| vote average type | period | auditor/treas incl | mean Dem vote share |
|---|---|---|---|
| *Average over elections* | | | |
| | 2012-20 | yes | 52.85% |
| | 2012-20 | no | 53.48% |
| | 2016-20 | no | 52.65% |
| *Vote index average* | | | |
| | 2012-20 | yes | 52.59% |
| | 2012-20 | no | 53.20% |
| | 2016-20 | no | 52.36% |

The expert witness testimony covering 2012-20 including the auditor general and treasurer races yields an average democratic vote share within 0.2% of the average vote share for 2016-20 *not* including auditor general and treasurer races. Thus the 2016-20 sample *not* including auditor general & treasurer is a close proxy for the 2012-20 sample *including* auditor general & treasurer that we use throughout this analysis. If anything, the 2016-20 sample without auditor & treasurer races show slightly more pro-Republican metrics than the 2012-20 sample including those two races.

Plan Baselines

In the tables we show base metrics for plans under contention and subject of expert witness testimony, first for the PA House. For completeness we include both the PA Legislative Reapportionment Commission (LRC) preliminary and final plans, with minor differences between them, together with an amended map filed by PA House Majority Leader Kerry

Benninghoff. A primary purpose of this paper is comparing these baseline metrics to results from random ensembles, and exploring differences between the Gerrychain and Redist ensemble methods. In these baseline maps we consolidated all split precincts so we could analyze them with Gerrychain and Redist thus small differences in metrics exist compared to tabulations using census blocks as a base unit.

Leader Benninghoff's stated concern in his proposed amendment was insufficient minority population in opportunity districts to elect minority representatives of choice however main difference in his plan metrics compared to the LRC maps is fewer Democratic seats and a more pro-Republican efficiency gap (Stephanopoulos and McGhee 2015)[13,14]. Comparing our baseline estimates using the 2012-20 vs 2016-20 election samples, for the LRC maps the 2016-20 election sample results in fewer Democratic seats won, and stronger Republican biases in efficiency gap mean-median metrics, as shown in Table 3. Conversely, for the Benninghoff plan the less Democratic sample 2016-20 election sample yields more Democratic results. This is surprisingly pronounced for metrics tallied using the "sum of votes" index, where for the 2012-

---

[13] The efficiency gap complements the mean-median metric for votes tabulated by district assessing which party "wastes" more votes. A vote is "wasted" if a district is carried by more than a one-vote majority- all the extra votes are deemed wasted. If a district is lost then all the losing votes are "wasted." Stephanopoulos and McGhee (2015) defined efficiency gap (EG):

$$EG = \frac{WV_2 - WV_1}{total\_votes}$$

where the wasted votes per district for either party (1 or 2) in a two-party system are:
$$WV_{loser}(district) = votes(for\ loser, district)$$
$$WV_{winner}(district) = votes(for\ winner, district) - votes(total, district)/2$$

[14] Efficiency gap in Redist > 0 for pro-Republican values, opposite to values in Gerrychain functions, it is also opposite to the sign convention in the Barber testimony. Here our convention for efficiency gap and mean-median is *positive* for *pro-Republican* values. We adjust Gerrychain results and testimony accordingly.

20 sample Democrats won 53.2% of the statewide vote yet only win 99 seats/ 203, representing 48.8% of the seats.

Table 3: Baseline Metrics for Proposed PA House Plans

| PA House Plan | election sample | Dem Seats won | efficiency gap | mean-median | Polsby-Popper | vote index wins |
|---|---|---|---|---|---|---|
| LRC Prelim | 2012-16 | 108.36 | 0.0402 | 0.0188 | 0.3382 | 107 |
| LRC Prelim | 2016-20 | 107.25 | 0.0376 | 0.009 | 0.3382 | 106 |
| LRC final | 2012-16 | 108.54 | 0.0405 | 0.019 | 0.3388 | 107 |
| LRC final | 2016-20 | 107.38 | 0.0381 | 0.0107 | 0.3388 | 106 |
| | | | | | | |
| Benninghoff | 2012-16 | 104.45 | 0.05946 | 0.0366 | 0.3421 | 99 |
| Benninghoff | 2016-20 | 105.375 | 0.04522 | 0.0316 | 0.3421 | 104 |

Expert witness testimony on these example maps is mostly consistent with these results. Both expert witnesses used a 2012-20 election sample including treasurer & auditor general contests, nearly equal to our 2016-20 sample with respect to statewide Democratic voteshare (0.2% higher than our 2016-20 sample). The Republican (Barber testimony) measures 107 Democratic seats won using the vote index vs 106 in our sample for the LRC preliminary map; the Democratic (Imai testimony) measures 106 seats vs 107.25 in our sample.

The PA Congressional maps show a similar variation of metrics, where there are small reductions in Democratic seats won in the 2016-20 sample vs the 2012-20 sample. There is a significant difference between the adopted map (the Carter map, drawn by expert witness Jonathan Rodden as a "least change" version of the previous 2020 map) and the HB2146 map passed by the Legislature as a House Bill but vetoed by the Governor.

Table 4 shows the HB2146 Congressional plan tilts more pro-Republican than the enacted

Carter map. The 2016-20 vote sample with a lower Democratic statewide voteshare results in

low Democratic seats won and other metrics compared to the 2012-20 sample. Only the

Republican expert witness (Barber) provided testimony on seat wins using ensemble simulations, and his estimate of vote index seat wins (9) in the HB2146 equals our calculation.

Table 4: Baseline Metrics for PA Congressional Plans

| PA Congressional Plan | election sample | Dem Seats won | Fractional Dem | efficiency gap | mean-media | Polsby-Popper | vote index wins |
|---|---|---|---|---|---|---|---|
| Carter (enacted) | 2012-16 | 9.636 | 9.284 | -0.01074 | 0.0126 | 0.309 | 10 |
| Carter (enacted) | 2016-20 | 9.5 | 9.1735 | -0.0095 | 0.0066 | 0.309 | 10 |
| | | | | | | | |
| HB2146 (republican) | 2012-16 | 8.818 | 8.68 | 0.0372 | 0.0234 | 0.3096 | 9 |
| HB2146 (republican) | 2016-20 | 8.375 | 8.4486 | 0.0562 | 0.0181 | 0.3096 | 9 |

County & Municipal Constraints

Legal plans for Pennsylvania House districts restrict county and municipal splits.[15] A 2018

decision by the PA Supreme Court[16] construes similar criteria for Congressional Districts.

Therefore, in addition to unconstrained simulations of Congressional and Legislative Districts

we incorporated county and municipal splitting restrictions into runs using Gerrychain and

Redist.

---

[15]PA Constitution II §16 "unless absolutely necessary, no county or [municipality] shall be divided in forming a … district"). https://www.legis.state.pa.us/WU01/LI/LI/CT/HTM/00/00.HTM

[16] LWV vs. State of Pennsylvania  https://cases.justia.com/pennsylvania/supreme-court/2018-159-mm-2017-2.pdf?ts=1518041023

Within Gerrychain we incorporated a simple rejection constraint preventing any proposal with more that a critical maximum number of county & municipal splits from being accepted as a valid plan (see Appendix for technical details on constraint implementation)[. This was very inefficient, as per the speed tests, with only a few percent of plans surviving the limiting criteria. However, this constraint is transparent.

Autocorrelations in Congressional District Simulations:

In Gerrychain we implemented constraints using their Boltzmann factor framework with various weights. Not every method worked for each case- further details in the Appendix. In Figure 5 we show the autocorrelation of Democratic seats won for Gerrychain & Redist with county and municipal split constraints (a & b) then Redist with county split constraints alone (c.) for Congressional Districts.

Figure 5: Constrained Autocorrelation Functions

a. Gerrychain ACF County/ Muni Constraint         b. Redist ACF County/ Muni Constraint

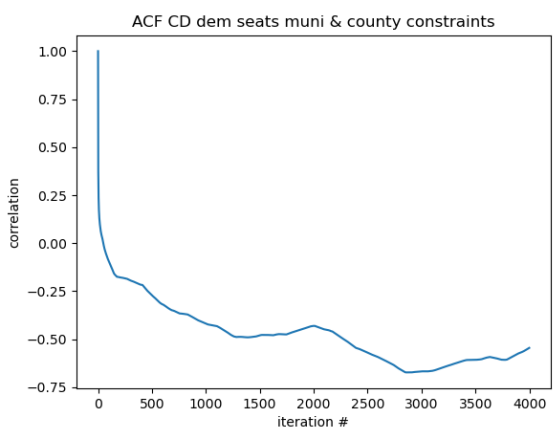

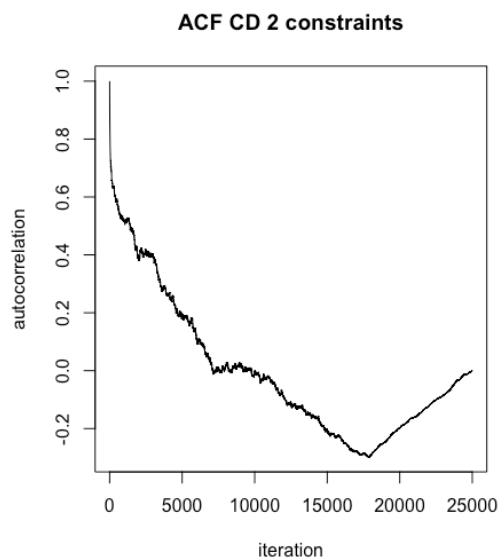

c. Redist ACF County Constraint Alone

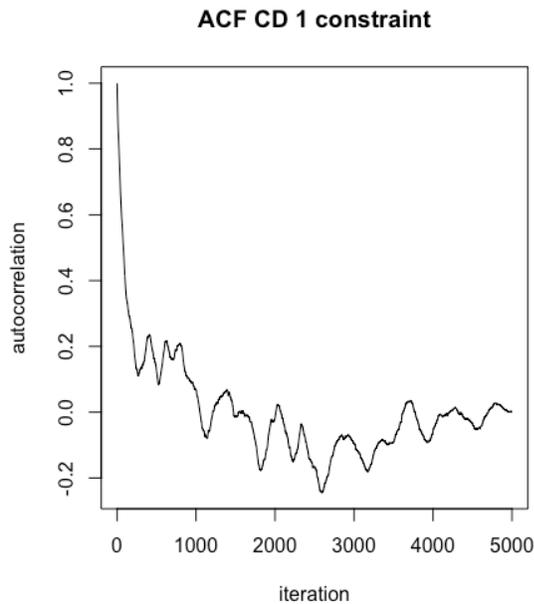

For two split constraints (county & municipal split constraints, plus multi-split constraints in Redist), both Gerrychain and Redist merge-split show that limiting the space of possible maps has an impact on the autocorrelation function: instead of declining to zero fairly rapidly and staying there, the autocorrelation function has negative overshoot. Figure 5 c. shows that considering county splits alone, the autocorrelation function remains close to zero after an initial sharp decline, suggesting better state mixing. With 2570 unique municipalities vs 67 unique counties in Pennsylvania, municipal splitting restrictions are far stronger than county restrictions. Given the presence of some correlation at long iteration values, we suggest the Markov chain simulation results for constrained Congressional Districts are indicative only, validated against unconstrained simulations and random tree simulations where autocorrelation is not present.

We found analogous results for constrained ensembles in PA House District simulations,  with Figure 6 showing autocorrelations for Redist and Gerrychain. Again, the autocorrelation functions oscillate after sharp declines from the first iterations, with the decline being more rapid and stable for the case where only county splits were constrained. Autocorrelation functions for ensembles with a single county constraint are better behaved. Here, again, we view the simulation results for the county and municipal constrained cases as being indicative, with validation from unconstrained and random tree cases helpful.

Figure 6: Autocorrelations in House District Simulations

a. Gerrychain ACF County/ Muni Constraint                  b. Redist ACF County/ Muni Constraint

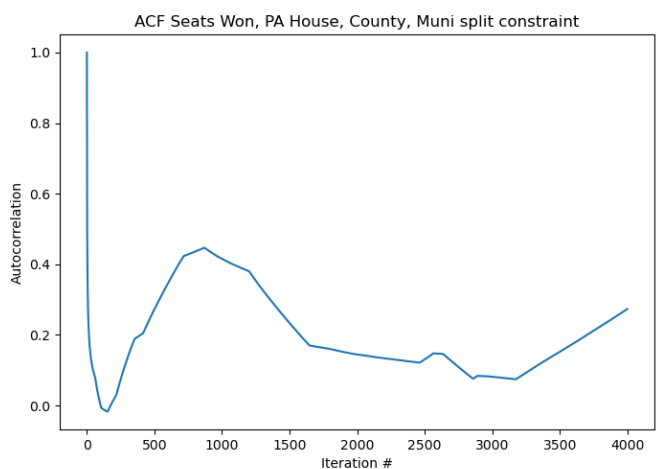

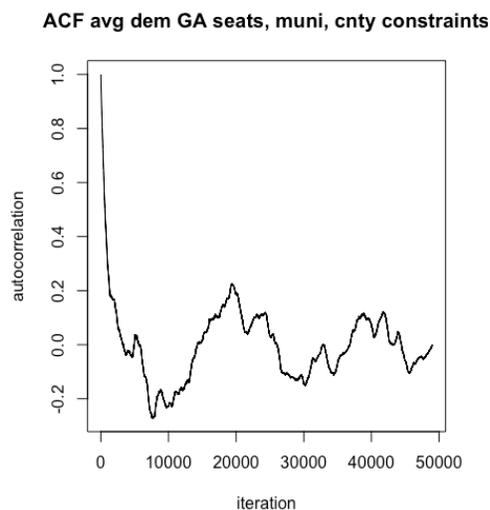

c. Gerrychain ACF County Constraint Alone

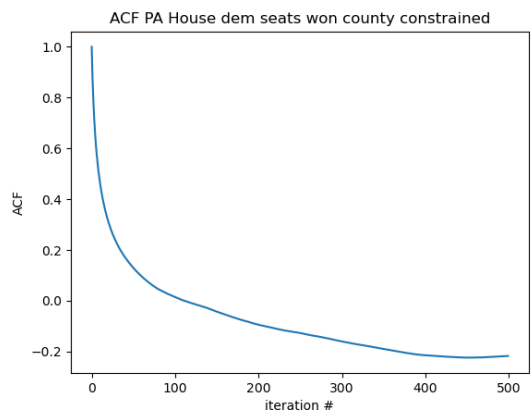

Ensemble Results

We describe ensemble results using the same metrics as in the expert witness testimony:

democratic Congressional (CD) and House District seats won together with efficiency gap and mean-median. For compactness measure we show the Polsby-Popper metric data. Results for democratic seats won differ depending on the election samples used.

We refer to the "fractional win" concept for close elections (Nagle and Ramsay 2021). For

example, when a party has 50.00001% vote share in a district they are not credited with an

outright seat win but rather 0.5 seats. For results near 50% the seat allocation is smoothed out

using a probability distribution function using a 5% standard deviation in voteshare over similar

elections For Pennsylvania. While fractional seats are not too important for the 203-seat PA

House, for Congressional seats they give different results from outright seat wins.

We compare Markov chain evolution for index, composite and fractional seat wins in Figure 7:

Figure 7: Markov Chain Evolution of Different Seat Win Metrics

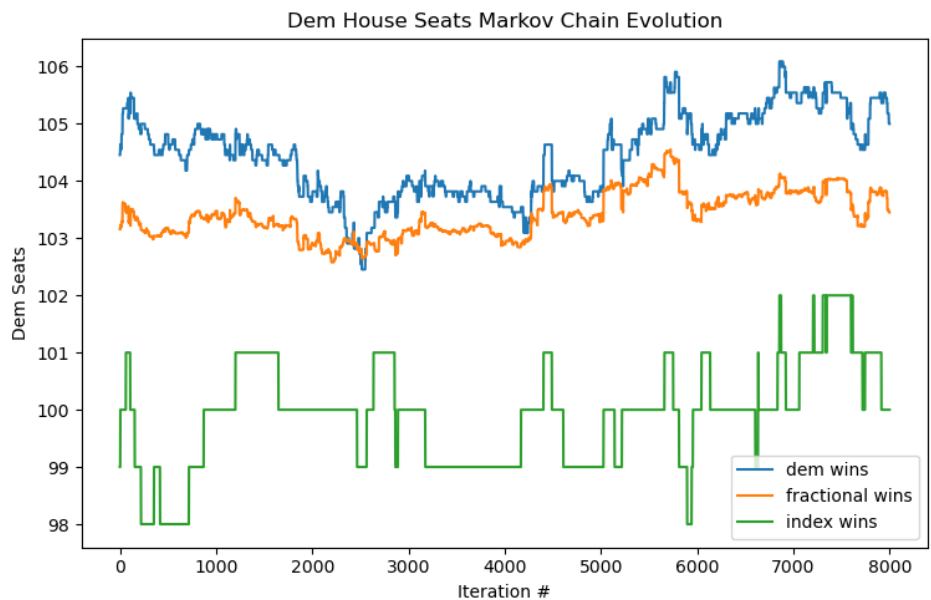

Because index wins assess the number of seats won once per map, for a single vote total, its values are quantized over integer seat numbers as opposed to seat wins where each election is calculated separately per map to obtain seat wins that are then averaged over all the elections in a sample. Index wins thus don't evolve as continuously. Vote indices also assign more weight to races taking place in presidential election years, when voter turnouts are far higher, notwithstanding the purpose of ensembles for estimating metrics for races that occur equally every two years.

Considering the indicative nature of the county & municipal split constrained simulations using Redist's merge-split or Gerrychain's Recom suggested by their autocorrelation functions, we compare to ensemble methods without map sampling issues. For Congressional districts Redist's SMC algorithm is well-behaved even if constrained for county & municipal splits. For both Congressional and House districts we show results with Gerrychain's random tree method. While this method doesn't permit incorporating county or municipal split constraints it provides a data point we tie to constrained results by progressively relaxing split constraints for Gerrychain's Recom method.

PA House

Figure 8 compares Gerrychain's Recom ensemble with 8000 iterations x 40 processors to the random tree ensemble on the 2016-2020 election data sample using 200 iterations x 40 processors. Here, the Recom simulation is illustrated in blue, the random tree results are in green, and the seats won in the enacted final plan indicated by the vertical black line. The initial seed for the Recom run was the PA Legislative Districting Commission final plan.

Figure 8: PA House: Gerrychain Recom & Random Tree Compared, Democratic Seats Won

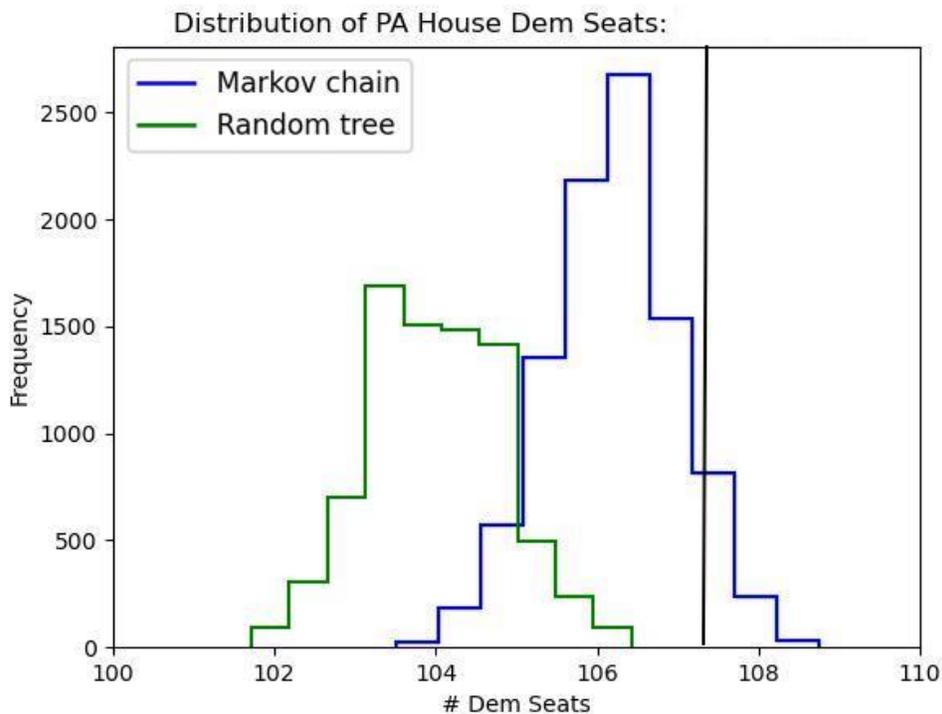

Similarly, we performed a constrained Redist merge-split run with the same sample (2016-2020 elections, initial seed plan the PA LRC preliminary map), and 50,000 single-processor chain iterations- depicted in Figure 9. We used a custom county and municipal split constraint, described in Appendix. Results are similar but approximately two fewer democratic seats as compared to the Gerrychain Recom simulation of Figure 8.

Figure 9: PA House: Redist merge-split, democratic seats won, blue showing showing distribution mean

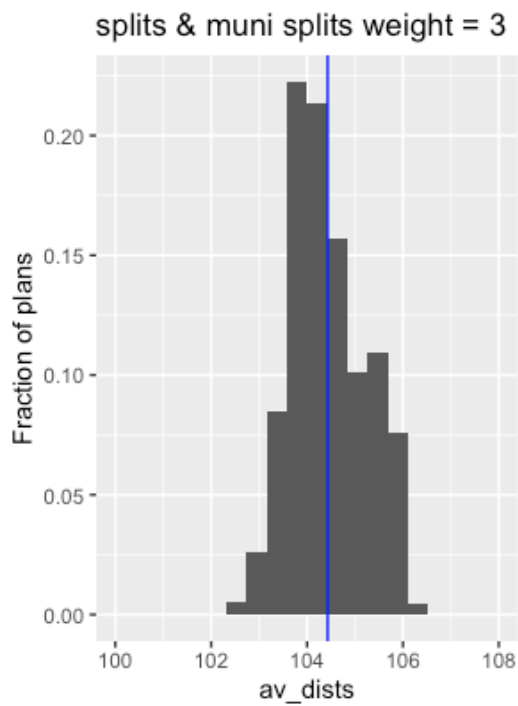

We performed Gerrychain and Redist runs over a variety of run conditions. To test the dependence of model results on an initial seed we compared Markov chain runs started using either the Legislative Reapportionment Commission's plan to the amended plan filed by Pennsylvania House Majority Leader Kerry Benninghoff. We compared different election data samples (the 2012-2020 and a subset from 2016-2020). Finally, we performed validation runs with random trees omitting county or municipal splitting constraints.

In Figure 10 we show how mean democratic seats won in Gerrychain ensembles scales with increasing county splits, indicating that the Recom model converges to the same result as the random tree method as county/ municipal constraints phase out (red dot indicates random tree ensemble, blue dots indicate Markov chain ensembles). The random tree method yielded democratic seats won of 103.53 and 103.33 for the 2012-20 and 2016-20 election sample respectively. Relaxing county/ municipal boundary splitting constraints improves the autocorrelation function for the Markov chain, yielding 102 and 106 democratic seats won for the 2012-20 and 2016-20 election samples. The best fit line to the Markov chain runs extrapolates to a value of Democratic seats won 0.4 higher than the random tree case at high county splits. This supports some validity of the Recom model despite the presence of autocorrelation.

Figure 10: PA House Scaling of Seats Won with County Split Constraint Strength

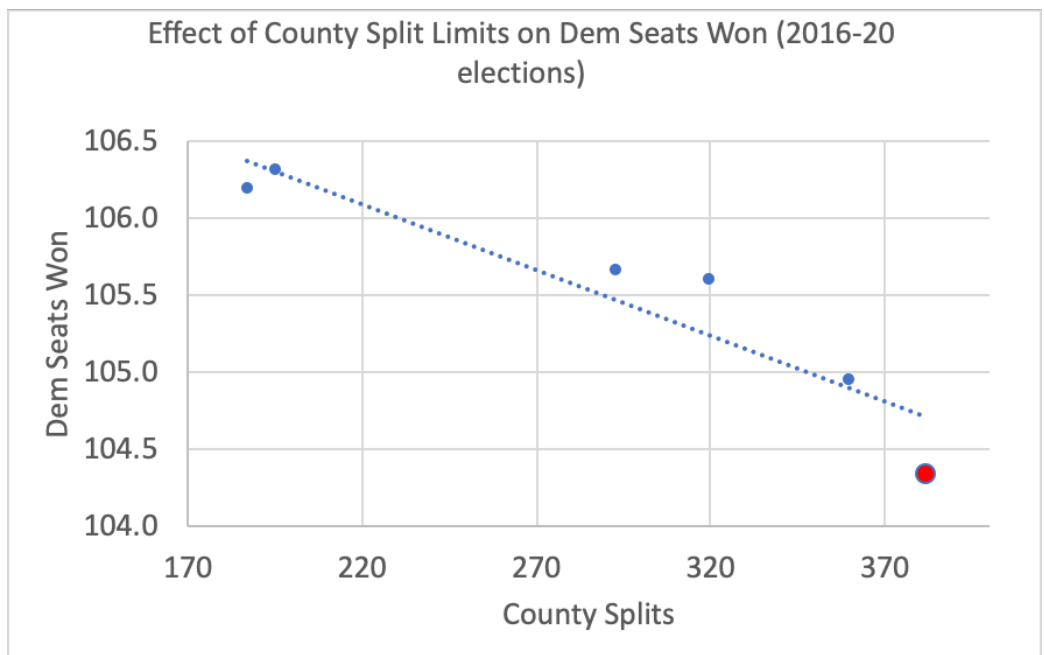

The metrics for constrained runs allow a direct comparison between Gerrychain (Table 5) and Redist (Table 6) using the same recombination algorithm (Recom vs merge-split). Changing the election sample [from 2012-20 to 2016-20 reduces Democratic seats won by about 1 seat for Gerrychain and 3 seats for Redist. There was a systematic shift towards republican-leaning values (seats won, efficiency gap and mean-median metrics) when the initial seed was the Benninghoff amended plan vs the Committee plans. A 2.5 democrat seat won difference using Gerrychain between the two initial seeds compares to a 1.5 democrat seat won difference with Redist merge-split, suggesting both methods sample the space of feasible maps unevenly but not grossly.

Table 5: Gerrychain Run Summary Metrics- PA House

| gerrychain simulation type | election sample | initial state | county splits | muni splits | mean seats won | +/- std dev seats | vote index/ seats won | +/- std dev seats | polsby-popper | efficiency gap | mean-median |
|---|---|---|---|---|---|---|---|---|---|---|---|
| markov chain | 2012-20 | LRC final | 195 | 107 | 107.15 | 0.9 | 104.26 | 1.7 | 0.320 | 0.044 | 0.027 |
| markov chain | 2016-20 | LRC final | 195 | 106.9 | 106.20 | 0.8 | 104.67 | 1.5 | 0.321 | 0.041 | 0.021 |
| markov chain | 2012-20 | Benninghoff | 195 | 106.9 | 104.67 | 0.8 | 99.80 | 1.6 | 0.321 | 0.056 | 0.034 |
| markov chain | 2012-20 | Benninghoff | 195 | 104.6 | 104.62 | 0.9 | 100.00 | 1.9 | 0.321 | 0.057 | 0.034 |
| markov chain | 2016-20 | Benninghoff | 195 | 106.9 | 105.10 | 0.9 | 104.20 | 2.1 | 0.321 | 0.045 | 0.026 |
| random tree | 2012-20 | N/A | 371.6 | 582.5 | 104.53 | 1.0 | 102.49 | 2.4 | 0.214 | 0.053 | 0.031 |
| random tree | 2016-20 | N/A | 382 | 586.00 | 104.33 | 1.2 | 103.215 | 2.14 | 0.215 | 0.045 | 0.025 |

Table 6: Redist Run Summary Metrics- PA House

| Redist simulation type | election sample | initial state | county splits | dem seats won mean | stdev | dem seats vote index | stdev | efficiency gap | mean-median | efficiency gap- index | mean-mean index |
|---|---|---|---|---|---|---|---|---|---|---|---|
| merge-split, 2 custom constraints | 2012-2020 | LRC prelim | 188.7 | 107.87 | 0.68 | 104.1 | 1.4 | 0.04 | 0.0224 | 0.0383 | 0.0181 |
| merge-split, 2 custom constraints | 2016-2020 | LRC prelim | 188.7 | 104.45 | 0.56 | 103.73 | 1.2 | 0.03832 | 0.0147 | 0.0339 | 0.012 |
| merge-split, 2 custom constraints | 2012-20 | LRC prelim | 188.9 | 108.95 | 0.434 | 103.6 | 1.2 | 0.392 | 0.24 | 0.0495 | 0.0186 |
| merge-split, 2 custom constraints | 2016-20 | LRC prelim | 188.9 | 105.07 | 0.48 | 102.97 | 0.75 | 0.386 | 0.179 | 0.0452 | 0.0113 |
| merge-split, 2 custom constraints | 2012-20 | bennninghoff | 190.7 | 106.37 | 0.8 | 100.9 | 2 | 0.06058 | 0.0362 | 0.0811 | 0.0385 |
| merge-split, 2 custom constraints | 2016-20 | bennninghoff | 190.7 | 103.38 | 0.78 | 100.07 | 1 | 0.0498 | 0.0311 | 0.0657 | 0.029 |
| merge-split, NO CONSTRAINTS | 2012-20 | LRC prelim | 374.6 | 106.37 | 1.13 | 103.82 | 2.8 | 0.0408 | 0.0189 | 0.0406 | 0.0083 |
| merge-split, NO CONSTRAINTS | 2016-20 | LRC prelim | 374.6 | 102.13 | 1.24 | 102.14 | 2.9 | 0.037 | 0.01 | 0.0362 | 0.0093 |

PA Congressional

Similarly, we compare the Gerrychain Markov chain data using Gerrychain's Recom method to the random tree method in Figure 11, using 2016-2020 election data. Here, the Recom simulation is illustrated in blue and the random tree in green. The value of the enacted Carter plan is the vertical black line. The initial seed for the Recom run was the enacted Congressional plan. Although the random tree plan and county & municipal-split constrained Markov chain distributions overlap, they differ in part because of the lack of split constraints in the random tree model.

Figure 11: Distributions of Congressional Seats Won in Markov Chain vs Random Tree

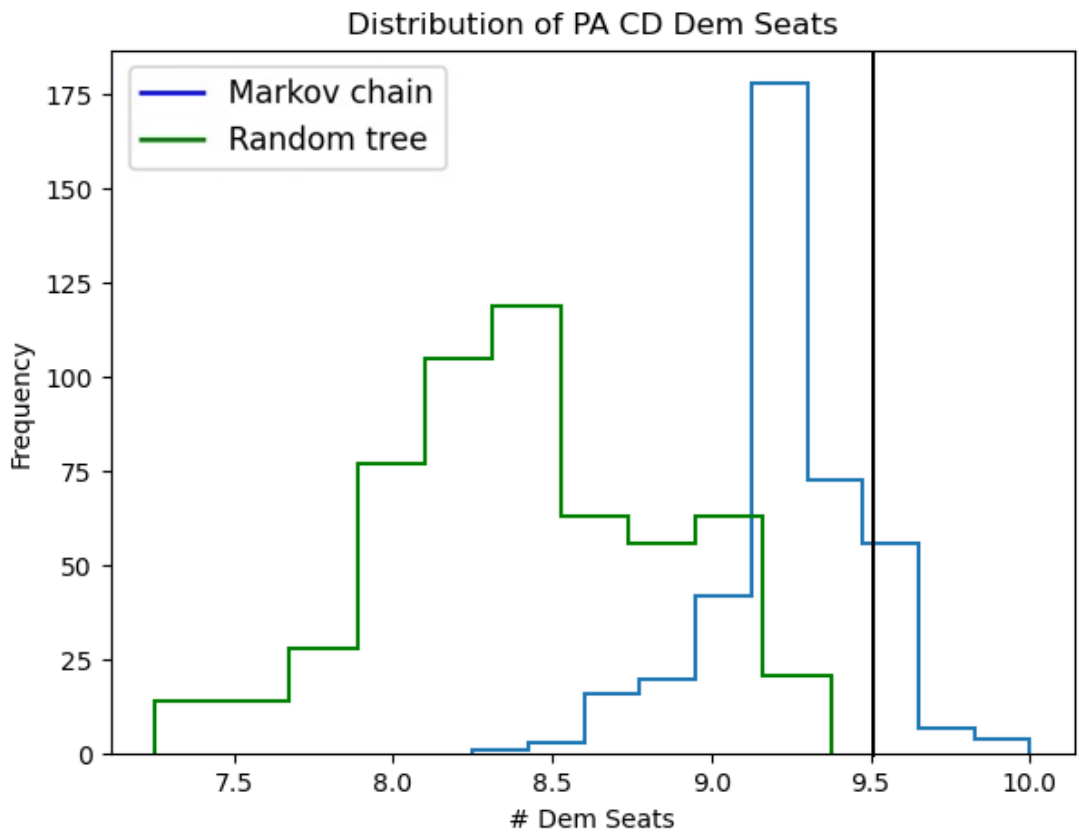

As with our analysis for Statehouse districts, we analyzed dependence on county split constraints for the Markov chain model and see that as split constraints are relaxed, mean Democratic congressional seat wins approach the random tree model- shown in Figure 12. In these runs we sampled 2012-20 election data using the enacted Carter plan as the initial seed for Markov chain. The best fit line yields Democratic seat wins a few tenths of a seat higher than the random tree case.

Figure 12: Scaling of Congressional Seats Won with County Split Constraint Strength

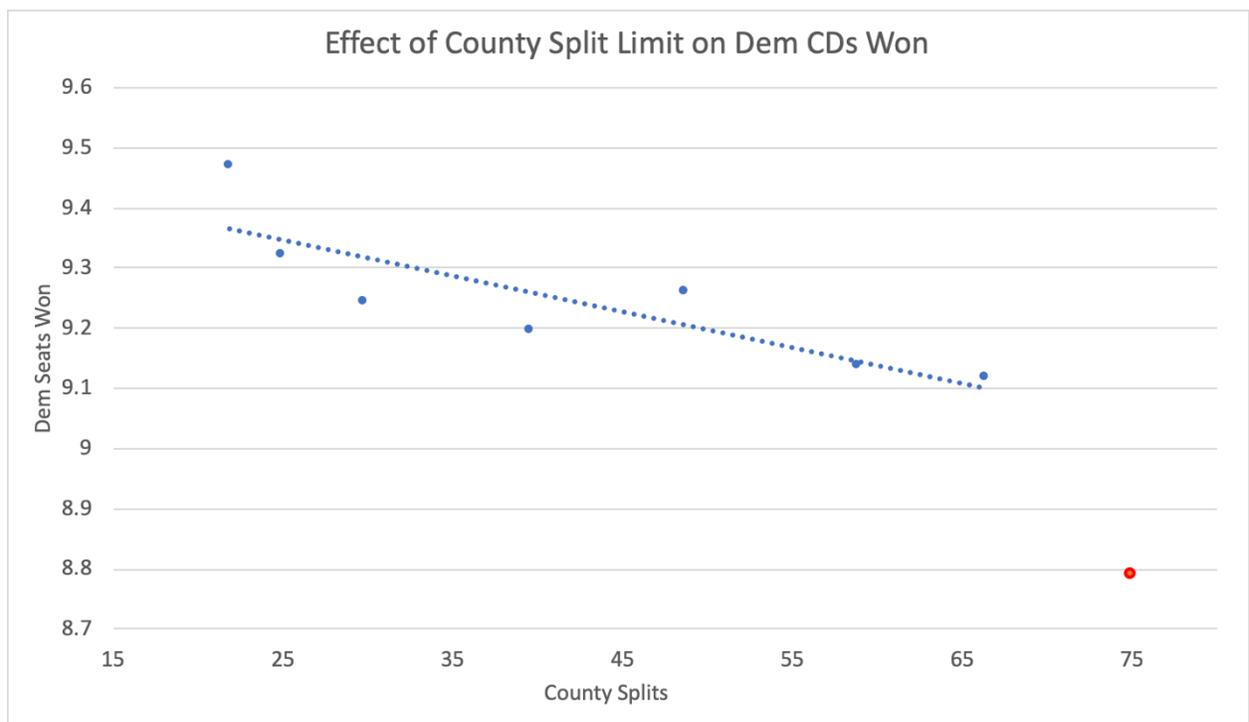

In Figure 12 the red dot indicates Democratic congressional seats won for the random tree case and blue indicates the Markov chain simulations with progressively relaxed limits on county split constraints.

Constrained Markov chain runs gave results shown in Table 7 for mean Democratic seats won between 9.22 and 9.69 depending on run conditions using the enacted plan as an initial seed, however run results were between 8.65 and 9.02 Democratic seats using the House Republican plan (HBN2146) as an initial seed. In both cases, run results were lower by several 10th's of a seat using the 2016-20 election data sample as compared to the 2012-20 election data sample.

Table 7: Gerrychain Run Summary Metrics- PA Congressional

| gerrychain simulation type | election sample | initial state | county splits | muni split limit | mean seats won | +/- std dev seats | vote index/ seats won | +/- std dev seats | polsby-popper | efficiency gap | mean-median |
|---|---|---|---|---|---|---|---|---|---|---|---|
| Markov chain | 2012-20 | Carter | 22 | 50 | 9.693 | 0.19 | 9.99 | 0.09 | 0.2892 | -0.0149 | 0.0128 |
| Markov chain | 2012-20 | Carter | 22 | 50 | 9.472 | 0.23 | 9.9 | 0.3 | 0.2767 | -0.003 | 0.0138 |
| Markov chain | 2016-20 | Carter | 22 | 50 | 9.22 | 0.24 | 9.72 | 0.45 | 0.277 | 0.005 | 0.007 |
| Markov chain | 2012-20 | HB2146 | 22 | 40 | 9.02211 | 0.22 | 9.86 | 0.35 | 0.294 | 0.013 | 0.016 |
| Markov chain | 2012-20 | HB2146 | 22 | 60 | 8.6563 | 0.13 | 9.91775 | 0.274 | 0.28 | 0.0152 | 0.0158 |
| Markov chain | 2016-20 | HB2146 | 22 | 50 | 8.65 | 0.12 | 9.05 | 0.22 | 0.289 | 0.0417 | 0.0097 |
| | | | | | | | | | | | |
| random tree | 2012-20 | N/A | 75 | 157 | 8.79 | 0.3 | 8.89 | 0.9 | 0.158 | 0.03895 | 0.0311 |
| random tree | 2016-20 | N/A | 73.325 | 151 | 8.4 | 0.45 | 8.29 | 0.88 | 0.163 | 0.054 | 0.0307 |

Redist's merge-split algorithm fails to sample a space of plans with both county and municipal split constraints, indicated by its slowly decaying autocorrelation function, thus there is no direct comparison to Gerrychain's Recom. Instead we rely on Redist's SMC algorithm, with quantitative metrics shown in Table 8. For the Congressional map unlike the State House, SMC converges. Using SMC while sampling 2012-20 elections we obtain the histogram shown in Figure 13 for Democratic congressional seats won, where the vertical line represents the mean of the distribution.

Figure 13: Redist SMC Ensemble Distribution of PA Congressional Seat Won

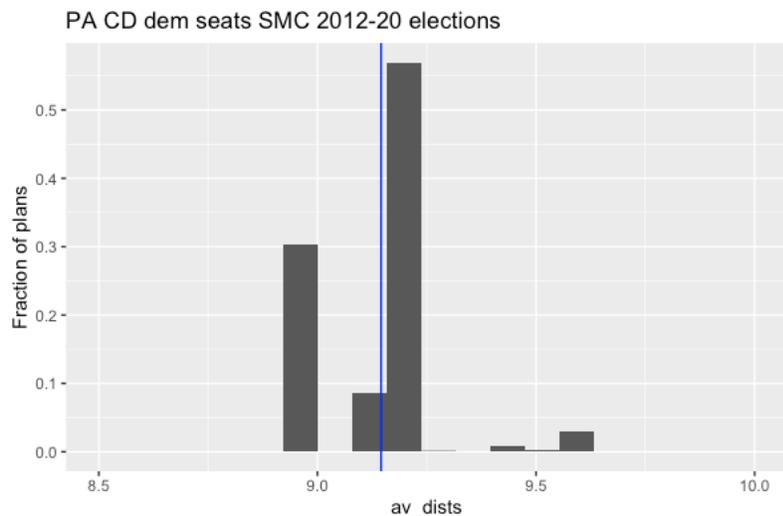

Table 8: Redist Run Summary Metrics- PA Congressional

| Redist simulation type | election sample | dem seats won mean | stdev | dem seats vote index | stdev | efficiency gap | mean-median | efficiency gap- index | mean-median index |
|---|---|---|---|---|---|---|---|---|---|
| SMC | 2012-2020 | 9.22 | 0.15 | 8.37 | 0.07 | -0.0107 | 0.01265 | -0.0386 | 0.00656 |
| SMC | 2016-2020 | 8.71 | 0.27 | 8.33 | 1.2 | -0.00637 | 0.005822 | -0.054 | 0.00744 |

We view a convergent SMC result as a standard because it doesn't depend on an initial state, as does a Markov chain, and because all its maps are uncorrelated (unlike a Markov chain). Its values for Democratic seats won are intermediate between Gerrychain's results for HB2146 and Carter map initial seeds.

The Redist composite vote index results are far below the results using Gerrychain regardless of initial seed. The efficiency gap metric (-0.054) based on vote index shows pro-*Democratic* bias: implausible considering Pennsylvania's Republican-friendly political geography where Democratic voters are mostly packed inefficiently in Philadelphia and Allegheny county. Using the vote index for metrics like efficiency gap makes little sense because efficiency gap is inherently non-linear: looking at a series of elections for a single seat, the wasted votes for a "sum of elections" does not equal the sum of the wasted votes per election. Nor does a sum of votes index appear quantitative for tallying seats won:

if there are 5 elections for a single seat that split 51/49% for parties A & B in 4 contests, then in the 5th election the result is 20/80% the average of the elections shows 4 seats won by party A and 1 seat won by party B; but summing all the votes gives the incorrect result that party B won 5 seats. If the turnout in the 5th election is higher, the vote index method favors party B even if the result for the 5[th] election is close.

Comparisons with Expert Witness Testimony

We compare results for PA House with simulations presented in the testimony by expert witness Michael Barber (on behalf of State Republicans)[17] and Kosuke Imai (on behalf of State Democrats)[18]. Imai skirts the convergence problem we faced with Redist/ SMC by breaking the State into 5 dense population centers modeled separately plus the remaining region. Barber presents results for the sum of votes index while Imai presents results for the average of elections. If these two methods were the same, their results would be obviously inconsistent, with Imai's distribution centered at 107 Democratic seats and not overlapping with Barber's result centered at 97 seats. Here we try to reconcile the dueling experts with our calculations of the Democratic seats won metric for both averages over elections and by sum of votes index.

---

[17] Written testimony by Michael Barber, Exhibit A:
https://www.pacourts.us/Storage/media/pdfs/20220218/152801-feb.17,2022-petitionforreview.pdf
[18] Written testimony by Koskue Imai, Exhibit B: https://www.pacourts.us/Storage/media/pdfs/20220311/181057-march11,2022-intervenor'sanswertopetitionforreview.pdf

We show the Democratic witness's histogram of Democratic House seats won, in Figure 14, and the Republican witness's histogram of Democratic House seats won using the vote index as a measure, in Figure 15.

Figure 14: Imai race-blind SMC model for PA Democratic House Seats Won

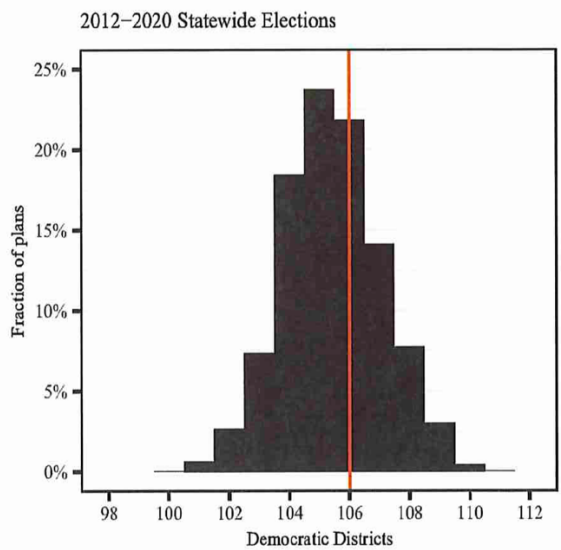

Figure 15: Barber Vote Index model for PA Democratic House Seats Won

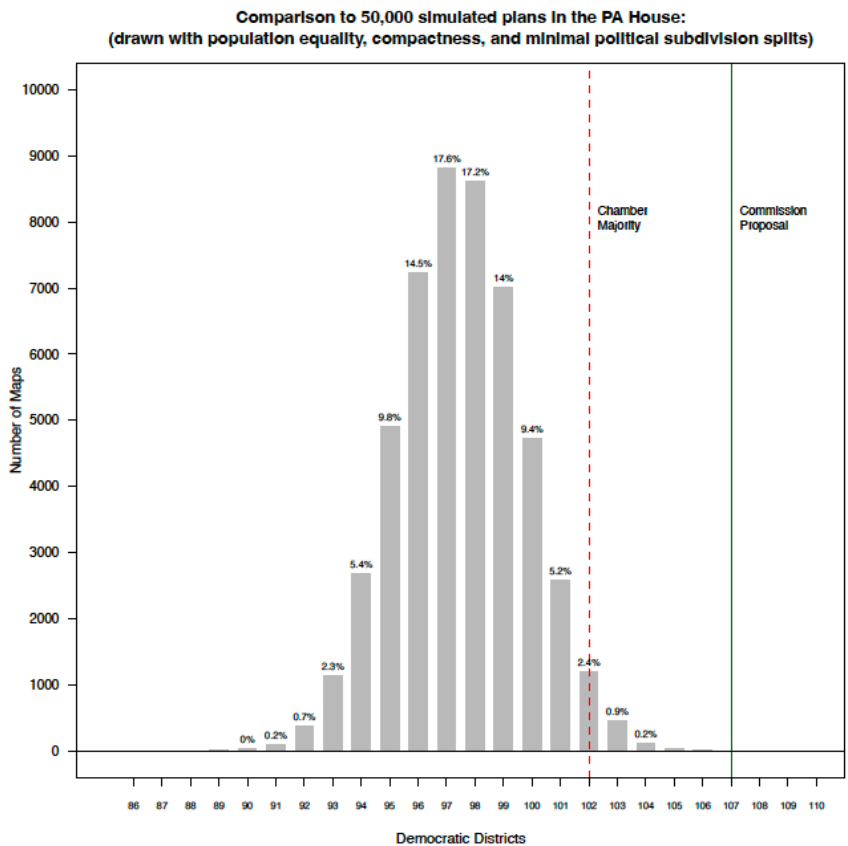

We use the distributional information (mean and standard deviation) to compare these expert witness reports to our Redist modeling (bars indicate the width of distributions from simulations not measurement errors). In Figure 16 we compare the Imai simulation (red point) to our Redist and Gerrychain modeling.

Figure 16: PA House Comparison of Democratic Expert Witness to Ensemble Modeling

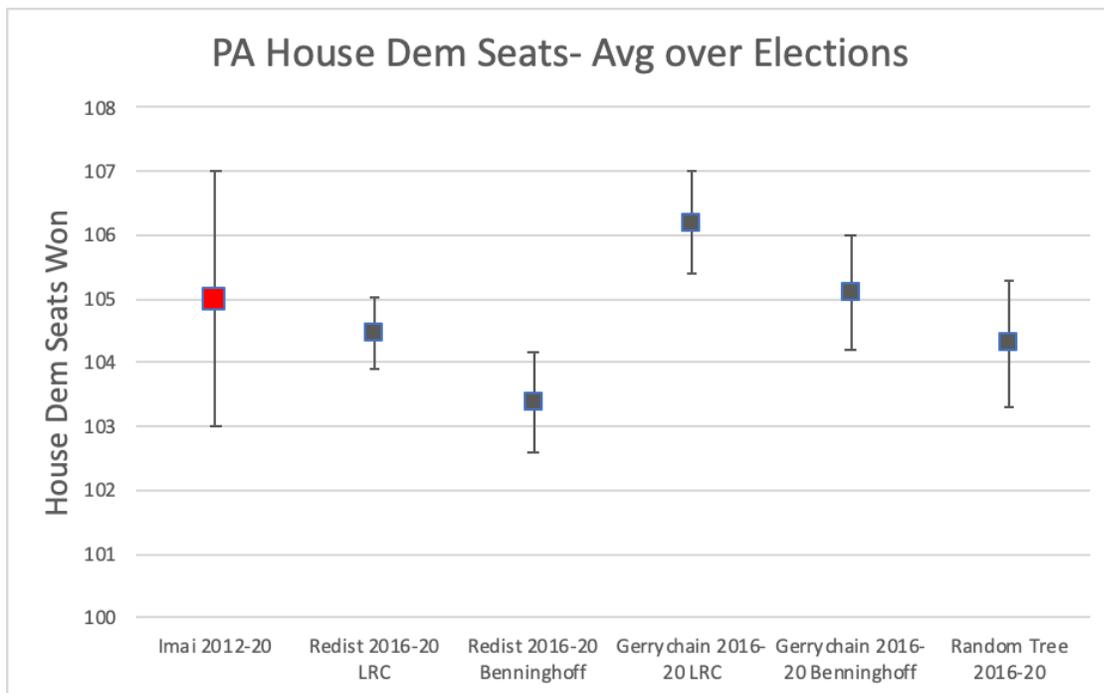

As discussed earlier, our election sample from 2016-20 is the best proxy for expert witness testimony covering 2012-20 elections including treasurer and auditor general races because the 2016-20 sample without the downballot races gives only 0.2% lower democratic statewide support. The Imai result is in the middle of the pack of our results (blue points) with our Redist yielding somewhat lower values for Democratic seats won than Gerrychain, and Markov chain simulations yielding lower seats when seeded with a more Republican-favorable initial plan notwithstanding exclusion of a correlation length (roughly) of initial Markov steps. For comparison the random tree, even though there are far more county and municipal splits than in a legal plan, yields result lower than the Markov chain simulations seeded with the enacted plan but still higher than the chain seeded with the Republican-favorable plan.

In figure 17 we compare the Barber testimony results (green point) to our simulations tallying Democratic seat won using the vote index measure (blue points). Here, the Barber result is lower than any of our estimates, including the unbiased random tree simulation result. Our results show the same pattern where Redist/ merge-split estimates are somewhat lower than Gerrychain, and chains seeded with Republican-favorable plans yield lower Democratic seats won than simulations with the enacted plan.

Figure 17: PA House Comparison of Republican Expert Witness to Ensemble Modeling using

Vote Index

We provide the same comparisons for simulations of the Congressional districts in expert testimony. Here, there were simulations by witness for Republicans Michael Barber[19] but no corresponding ensemble analyses presented by Democrats.

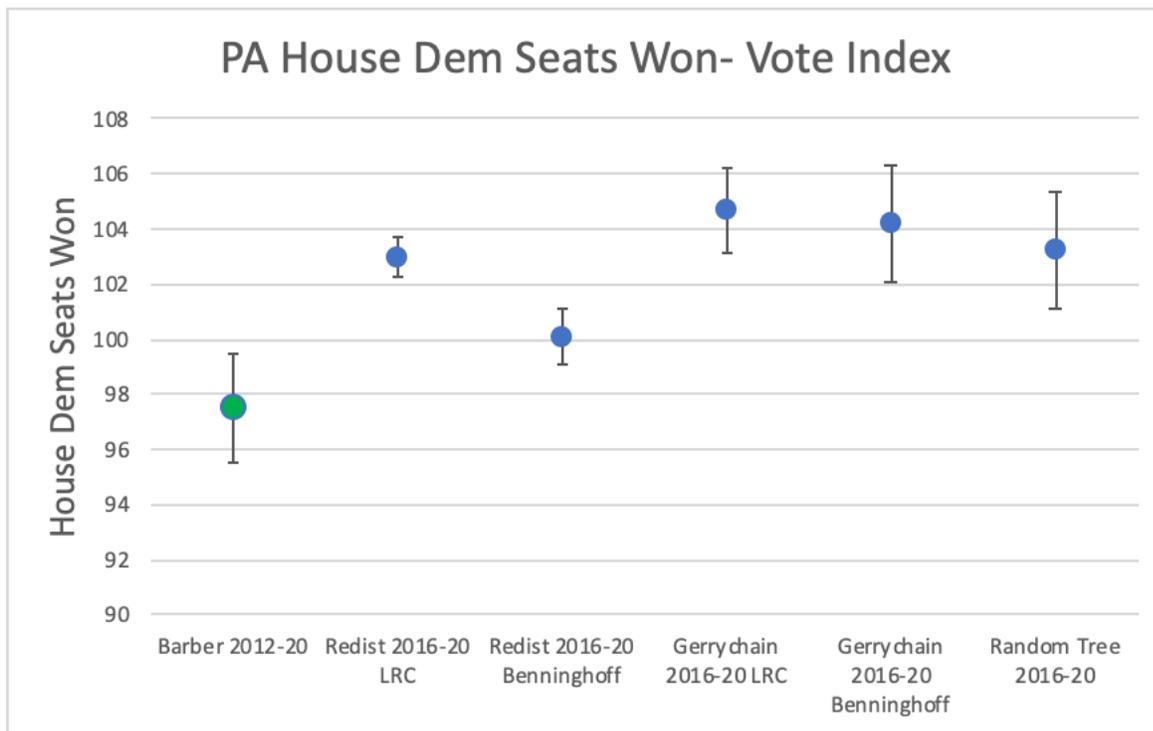

Drawing from the expert report, we estimate mean Democratic Congressional seats for the

Barber vote index of 8.4 and a standard deviation of one seat in the distribution (Figure 18).

Figure 18: Republican Expert Witness Vote Index Ensemble for Congressional Seats Wins

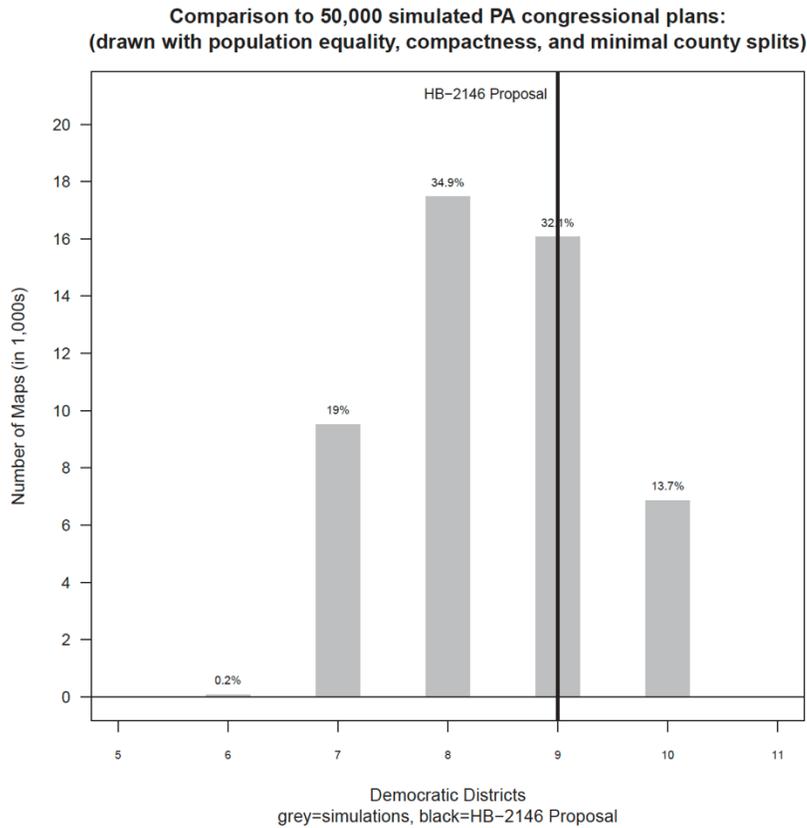

We compare to our results for the 2016 vote index in Figure 19 where the expert witness (green point) is consistent with the Redist SMC result for vote index and below the Gerrychain simulations while slightly above the random tree simulation without any constraints on county or municipal splits (blue points).

Figure 19: PA Congressional Seat Wins Comparison of Republican Expert Witness to Ensemble

Modeling using Vote Index

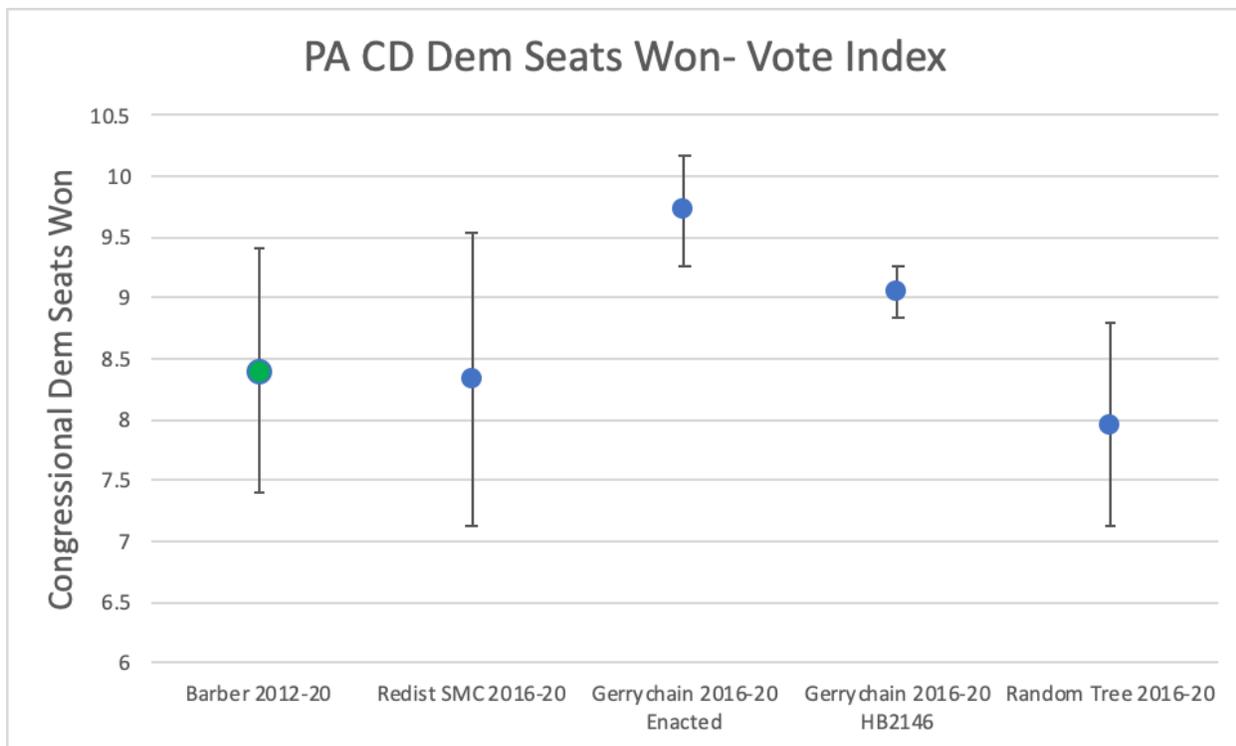

In Figure 20 we compare our results for Democratic seat wins, using the average number of

seats won over all elections, contrasting results with 2016-20 election sample (blue points) to

2012-20 sample (red points). Unsurprisingly, with average Democratic vote share of 52.65% vs

53.48% in the 2016-20 vs 2012-20 samples (not including auditor/ treasurer races), the latter results are higher across the board. The Redist SMC results are in the middle of the ranges calculated with Gerrychain for the most part. As with the State House seat simulations,

Gerrychain Markov chains using the enacted plan as an initial seed show higher Democratic seats than those with a Republican-favorable map. The county/municipal split unconstrained random tree method gives the lowest Democratic seat wins. Unlike for the PA House where average seat wins are far higher than seat wins computed using the sum of votes index, here the two metrics agree within the width of their distributions. But this is likely more coincidence than fundamental agreement because the methods differ so much.

Figure 20: Comparison of Ensemble Models for PA Congressional Seats Averaging over Elections

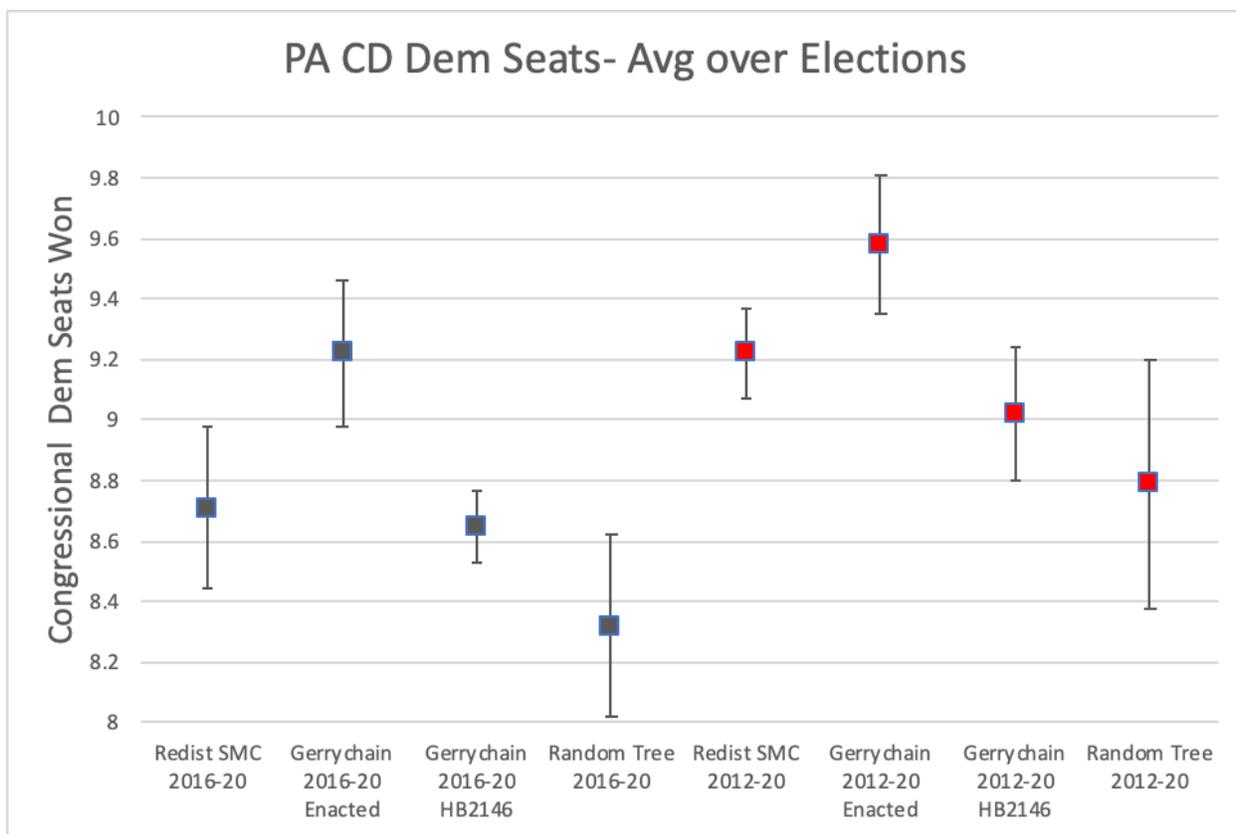

Conclusions

We used multiple methods to generate ensembles of PA State House and Congressional District maps: Tufts Gerrychain and Harvard's Redist. Within Gerrychain we used Markov chain and random tree models. Within Redist we used merge-split and sequential monte carlo (SMC). Using Redist, for the PA House merge-split allowed ensemble modeling with county and municipal constraints, and for PA Congressional districts only SMC produced ensemble distributions under these constraint conditions. We compared these results to the expert witness testimony presented in the PA House and Congressional District dockets.

Ensemble modeling with Redist was consistent with Gerrychain for metrics where we averaged over each election with equal weight. There was less consistency when metrics were calculated over a sum of votes index.

The Redist simulations are dramatically faster running than Gerrychain, with critical components are in C. However Redist's C code is harder to follow and customize.

Expert witness testimony using ensembles to confirm or refute the plausibility of proposed district plans prior to 2022 largely ignored constraints on county or other splits, so none of the ensemble elements bore any resemblance to a legal plan. More recent testimony using ensembles takes splits into account, particularly in Pennsylvania, a state that requires constraints on both county and municipal splits in legal plans. Because there are so many more municipalities in Pennsylvania (2570 municipalities vs 67 counties), municipal split limitations impose far more constraints on an ensemble.

If an ensemble is too highly constrained then it will not evenly sample the space of feasible plans. The sampling algorithm may only allow Markov chain steps exchanging the same precincts from one plan to a subsequent step then back again. This failure manifests itself in an autocorrelation function showing poor decay from an initial state, or even a state that refuses to evolve. We show autocorrelation functions for multiply-constrained ensembles for Gerrychain modeling PA House and Congressional districts. For Redist merge-split we show autocorrelation functions modeling PA House Districts but for PA Congressional districts the Redist merge-split did not evolve, suggesting some differences between the Redist merge-split algorithm and Gerrychain's Recom even though they are both ostensibly based on the same principles.

The autocorrelation functions for Gerrychain House districts & Congressional, Redist House districts show more limited decay within the first several hundred- 1000 Markov chain steps and subsequent oscillations than for a less constrained ensemble (eg. an unconstrained ensemble or an ensemble with only county split constraints). Furthermore, Gerrychain Recom and Redist merge-split both show some dependence on the initial seed state. This suggests the Markov chain ensembles may not be unbiased samples of the constrained map space. Therefore we extrapolated our these results with progressively relaxed county and municipal split constraints to a random tree result for state generation that does not depend on Markov state evolution. The approximate agreement of the unbiased random tree method to successively less constrained Markov chain ensembles provides some support the Markov chain methods provide an indicative sampling of the map space.

Redist SMC generates ensembles whose elements are inherently uncorrelated, an advantage over other methods suggesting it is an unbiased sample of the space of feasible maps. However Redist SMC method fails to converge for PA House ensembles. The Democratic expert witness, Kosuke Imai, circumvented this problem by dividing the State in six different regions and performing SMC models for each region separately although it is unclear whether mosaicking the regions together introduces artifacts. Nonetheless, the Imai result is consistent with our Redist merge-split and Gerrychain results.

One of the expert witnesses (Barber, for Republicans in both the House and Congressional district dockets) tallies a sum of votes index prior to computing metrics (seats won, efficiency gap, etc.) as opposed to computing metrics for each election separately then averaging each metric giving each election equal weight. We duplicated this vote index to validate this expert's work and found reasonable agreement for PA Congressional District ensembles. We could not replicate the Barber testimony for the PA House ensembles where he obtained results far more Republican-leaning than we could.

We found tallies using the vote index produced erratic results in some cases. For example, efficiency gap using vote index yielded strong pro-democratic bias that became even stronger for an electoral sample with a lower statewide democratic voteshare. Measures such as PA House Democratic seats won, when based on vote index tallies, fluctuated considerably more depending initial conditions as compared to averages over elections.

A possible reason is that seat wins using the vote index are smoothed out over many elections, as can be seen from Markov chain evolutions where no evolution occurs for many steps then exhibits a large jump.  Standard deviations for metrics calculated on sum of vote index are larger than if calculated on average of elections, suggesting data consolidation by the vote index introduces wide-tailed distributions of outcomes into the statistics.

Using the vote index weighs elections by the number of votes cast, thus is not an appropriate measure for estimating legislative seats won where presidential year and non-presidential year wins are equally important. Furthermore, metrics such as efficiency gap and mean-median are non-linear so summing over elections to create an index then computing a metric on the sum introduces distortions. We found the anomalous result that efficiency gap for PA Congressional district ensembles calculated in Redist-SMC using vote index is biased strongly pro-Democratic. This suggests the vote index is a flawed measure that should not be relied on, as others have already commented[20].

---

[20] Duchin, M. in NORTH CAROLINA LEAGUE OF CONSERVATION VOTERS, INC. et al vs Destin Hall in Superior Court of North Carolina (Feb 20 2022), 21 CVS 015426, 21 CVS 500085 pp. 13-15 "Faulty Averaging" and errors in weighting

Both expert witnesses in the Pennsylvania legislative proceedings used precinct-level election results including 2012-2020 contests that were not publicly available, namely the Treasurer and Auditor General races. We approximated the statewide partisan voteshares within 0.2% using 2012-2020 results from a public archive that did not include Treasurer and Auditor General races however expert witness should only use publicly available results where feasible to promote transparency.

Acknowledgements

We thank Cory McCartan and Christopher Kenny of Harvard University for their invaluable comments assisting us with Redist.

Appendix on Constraint Methods

Our code is in the Github repo https://github.com/dinosg/gerrychain_redist_compare.

We wrote functions in Gerrychain, using the documented updater method[21], counting administrative boundary splits. These are similar to the splits_count function in redistmetrics[22], except that redistmetrics counts the total number of district/ administrative unit pieces so that a whole county within a districts adds a count of 1 to splits_count. Thus the total number of counties in Pennsylvania (67) needs be subtracted from the redistmetrics splits_count function to determine total number of county splits.

---

[21] https://gerrychain.readthedocs.io/en/latest/api.html#module-gerrychain.updaters
[22] https://cran.r-project.org/web/packages/redistmetrics/redistmetrics.pdf

For Redist analyses of the PA House districts we used a custom constraint function because total_splits constraints failed to constrain splits. This custom constraint, eg.

```
add_constr_custom(constr,  0.3, # the constraint strength
   function(plan, distr) {
      check_ctys = unique(map$county[plan == distr]) # which counties this district touches
      is_split = sapply(check_ctys, function(cty) n_distinct(plan[map$county == cty]) > 1)
      sum(is_split) # penalty = # of counties split by this district
   })
```

works by counting the # of counties split by each district and applying a penalty to that as opposed to simply counting the total # of counties split.

Within Redist-SMC (for Congressional Districts) we were able to constrain county and municipal splits using the total_splits constraint function. These constraints are not logistic constraints but weighed using an exponential Boltzmann energy factor, so result in a distribution of county splits in the ensemble as opposed to an ensemble with split numbers concentrated on a maximum value specified as with Gerrychain.

For Redist merge-split and Gerrychain, adding constraints may reduce the Markov chain "mixing," that is the ability to sample independently shuffled states as seen by the evolution of the autocorrelation function.